# Evaluating Prompt Engineering Techniques for Accuracy and Confidence Elicitation in Medical LLMs

Nariman Naderi [1][0000-0003-4820-5497], Zahra Atf [2][0000-0003-0642-4341], Peter R Lewis [3][0000-0003-4271-8611], Aref Mahjoub far [4][0000-0002-1681-0994], Seyed Amir Ahmad Safavi-Naini [5][0000-0001-9295-9283], Ali Soroush [6][0000-0001-6900-5596]

[1] Division of Data-Driven and Digital Health (D3M), The Charles Bronfman Institute for Personalized Medicine, Icahn School of Medicine at Mount Sinai, New York, NY, USA
[2] Faculty of Business and Information Technology, Ontario Tech University, Oshawa, Canada
[3] Faculty of Business and Information Technology, Ontario Tech University, Oshawa, Canada
[4] School of Medicine, Iran University of Medical Sciences, Tehran, Iran
[5] Division of Data-Driven and Digital Health (D3M), The Charles Bronfman Institute for Personalized Medicine, Icahn School of Medicine at Mount Sinai, New York, NY, USA
[6] Division of Data-Driven and Digital Health (D3M), The Charles Bronfman Institute for Personalized Medicine, Icahn School of Medicine at Mount Sinai, New York, NY, USA and Henry D. Janowitz Division of Gastroenterology, Department of Medicine, Icahn School of Medicine at Mount Sinai, New York, NY, USA and The Charles Bronfman Institute of Personalized Medicine, Icahn School of Medicine at Mount Sinai, New York, NY 10029, New York, NY, USA

**Abstract.** This paper investigates the efficacy of prompt engineering techniques in enhancing both the accuracy and confidence elicitation of Large Language Models (LLMs) when applied to high-stakes medical contexts. A stratified dataset of Persian board certification exam questions, spanning multiple specialties, was used to systematically evaluate five LLMs— GPT-4o, o3-mini, Llama-3.3-70b, Llama-3.1-8b, and DeepSeek-v3. Each model underwent 156 unique configurations reflecting different temperature settings (0.3, 0.7, 1.0), prompt designs (e.g., Chain-of-Thought, Few-Shot, Emotional, Expert Mimicry), and confidence output scales (1–10, 1–100). The study employed metrics such as AUC-ROC, Brier Score, and Expected Calibration Error (ECE) to assess how accurately verbalized confidence matched real-world performance. Results revealed that while advanced prompting strategies—particularly Chain-of-Thought—consistently boosted accuracy, they also heightened overconfidence, indicating the need for post-hoc calibration. Emotional prompting inflated confidence further, potentially undermining clinical decision-making. Smaller models like Llama-3.1-8b exhibited marked underperformance across all metrics, emphasizing the importance of robust architectures in complex clinical scenarios. In contrast, proprietary models (e.g., GPT-based systems) demonstrated higher accuracy but still lacked reliable confidence calibration. These findings underscore the significance of designing prompts that effectively manage epistemic and aleatoric uncertainties, rather than solely focusing on accuracy gains. Ultimately, prompt engineering emerges as a dual-faceted approach—one that can substantially elevate model correctness yet inadvertently inflate confidence in erroneous outputs. Addressing this tension necessitates a combination of carefully crafted prompts and rigorous calibration protocols, especially where erroneous recommendations may have life-threatening consequences.

## 1      Introduction

Prompt engineering enhances the accuracy and confidence of large language models (LLMs) by guiding them to generate factually precise and well-calibrated responses [1][2]. In high-stakes applications, such as medical and scientific question-answering, robust prompt design is crucial for reducing hallucinations and ensuring reliability and interpretability [3][4][5]. Even minor variations in phrasing can significantly impact prediction confidence and uncertainty estimation [6]. Advanced techniques, such as Bayesian Prompt Ensembles, refine probability distributions using semantically equivalent prompts to improve uncertainty quantification—particularly in medical diagnostics—although maintaining semantic equivalence remains challenging [7][8][9][10]. The effectiveness of prompt engineering extends beyond accuracy enhancement, playing a pivotal role in the trustworthiness of LLM outputs, especially in applications requiring high precision and reliability.

The present research examines how different prompt engineering techniques influence LLM response accuracy and the calibration of confidence scores. It evaluates the effects of prompt phrasing on response reliability and explores methods to optimize prompt formulations to better represent both epistemic (knowledge-related) and aleatoric (inherent randomness) uncertainties, which is particularly important for improving model performance in medical applications.

Prior research demonstrates that prompt design significantly impacts model performance. Techniques such as self-consistency checks, probabilistic decoding, and structured prompting enhance response accuracy and confidence estimation [11]. Uncertainty-aware instruction tuning (UaIT) improves self-uncertainty representation, leading to better human-AI decision-making. Additionally, Chain-of-Thought (CoT) prompting breaks complex queries into logical steps to enhance reasoning [8], while Retrieval-Augmented Prompting (RAP) integrates external knowledge for greater accuracy—an essential feature in fields like medical diagnostics [6]. Furthermore, prompt perturbation consistency learning enforces consistency across query variations [11]. Despite these advancements, challenges persist in calibrating confidence levels and mitigating overconfidence, which can undermine trust in AI systems when users rely on seemingly confident yet inaccurate answers [12].

This investigation explores how different prompt engineering techniques affect both the accuracy and confidence calibration of LLM responses in the medical domain. It evaluates various prompt types—such as direct versus indirect queries, multi-shot prompting, and iterative refinements—to determine their impact on response precision and reliability in medical applications. Additionally, the study explores the integration of uncertainty estimation methods to enhance the representation of both epistemic and aleatoric uncertainties. By bridging prompt design and uncertainty estimation, this work provides practical insights into optimizing LLM interactions for more reliable, AI-assisted medical decision-making, addressing a critical need in clinical applications.

## 1.1   Contribution of This Study

This study systematically analyzes the impact of various prompt engineering techniques on accuracy and confidence elicitation in LLMs within the medical domain. The primary objective is to explore optimal strategies for improving the precision of model responses to medical queries and calibrating their confidence levels.

The key contributions of this study include the evaluation of prompt structuring techniques for medical queries, examining the influence of different prompt types—such as direct and indirect queries, multi-shot prompting, and iterative refinements—on the accuracy and reliability of model responses in medical applications. Additionally, it investigates the integration of uncertainty estimation methods in medical responses, analyzing the relationship between prompt design and confidence assessment to enhance the representation of epistemic and aleatoric uncertainty.

By bridging the gap between prompt design and uncertainty estimation, this study aims to provide actionable insights for practitioners seeking to deploy LLMs in sensitive applications like healthcare.

By providing a structured analysis of prompt engineering in medical question-answering, this study contributes to optimizing LLM interactions in healthcare. It aims to refine methodologies that improve response accuracy and confidence calibration, thereby fostering more reliable AI-assisted medical decision-making.

This work seeks to differentiate itself from existing research by offering practical solutions for balancing accuracy and confidence elicitation in medical contexts—a pressing need in the adoption of LLMs for clinical use.

## 2   Related Work

A growing body of work underscores the pivotal role of prompt engineering in enhancing both the accuracy and confidence elicitation capabilities of LLMs. Recent findings demonstrate that strategically structured prompts not only guide LLMs toward producing more contextually coherent and factual responses but also foster better self-awareness of uncertainty. Liu et al. [11] show that well-designed prompts can reduce hallucinations and factual errors while improving the interpretability of model confidence. Tonolini et al. [12] introduce Bayesian Prompt Ensembles, which combine multiple prompts via variational inference to enhance prediction reliability and calibration. Agrawa et al. [5] explore self-checking and direct question-answering strategies to further mitigate overconfidence and improve factual consistency.

Uncertainty quantification forms another key dimension of prompt engineering. Ling et al. [13] demonstrate that Bayesian inference techniques effectively capture both epistemic and aleatoric uncertainty, thereby enhancing model reliability in natural language understanding tasks. This finding is particularly relevant to prompt engineering, as different prompt formulations influence how an LLM expresses or withholds confidence. Xiong et al. [4] show that human-inspired prompting (e.g., CoT, Self-Probing) can improve calibration and accuracy, although no single method performs best across all tasks. Ailem et al. [14] reveal that prompt biases affect evaluations, while Yang et al. [9] demonstrate that rephrased queries—via synonym substitution and verbosity adjustments—yield more reliable uncertainty estimates.

Xu et al. [15] combine supervised fine-tuning and reinforcement learning to promote self-reflective rationales and reduce calibration errors, illustrating the synergy between prompt design and fine-tuning. Becker et al. [10] propose an explanation-stability framework using logical entailment to enhance confidence signals, while Errica et al. [7] introduce sensitivity and consistency metrics to evaluate prompt variations. Azimi et al. [8] demonstrate that tailored strategies, such as CoT-SC, improve reliability and accuracy in specialized tasks. Yadkori et al. [9] employ iterative prompting with an information-theoretic metric to separate epistemic from aleatoric uncertainties and detect hallucinations in multi-label tasks. Savage et al. [16] compare uncertainty proxies for medical diagnosis, finding sample consistency more effective than other metrics despite calibration issues. Wu et al. [17] show that Answer-Augmented Prompting can boost accuracy and consistency but may expose vulnerabilities, recommending fine-tuning and input prefix strategies to mitigate risks.

Collectively, these studies illustrate the far-reaching implications of prompt engineering for both accuracy and confidence in LLM outputs. By refining input formulations, experimenting with ensemble methods, adopting uncertainty quantification techniques, and iterating on prompt variations, researchers and practitioners can systematically reduce hallucinations, enhance calibration, and guide models toward more trustworthy behavior. As LLMs become increasingly integrated into real-world applications, prompt engineering stands out as a key methodology for advancing both performance and reliability.

## 3      Methodology

As shown in Figure 1, the methodology begins with curating a stratified dataset drawn from Persian medical board certification exam questions, ensuring a balanced and representative sample of clinical domains. This dataset feeds into an LLM pipeline configured to systematically vary temperature settings (three distinct levels where possible), prompt formulations (six different designs), confidence output formats (two scales), and model architectures (five separate LLMs). These configurations yield 156 unique runs, each capturing the model's answer choice (from four options), its explanation, and a corresponding confidence score (ranging from 1 to 10 or 1 to 100). The resultant outputs are subsequently evaluated along three dimensions: discrimination (via metrics such as receiver operating characteristic curve (AUC-ROC)), calibration (utilizing Brier scores, ECE, and calibration curves), and exploratory analyses (examining accuracy, mean confidence, and statistical significance).This layered approach illuminates how prompt engineering choices influence both the accuracy of generated responses and the calibration of confidence levels in a high-stakes medical context.

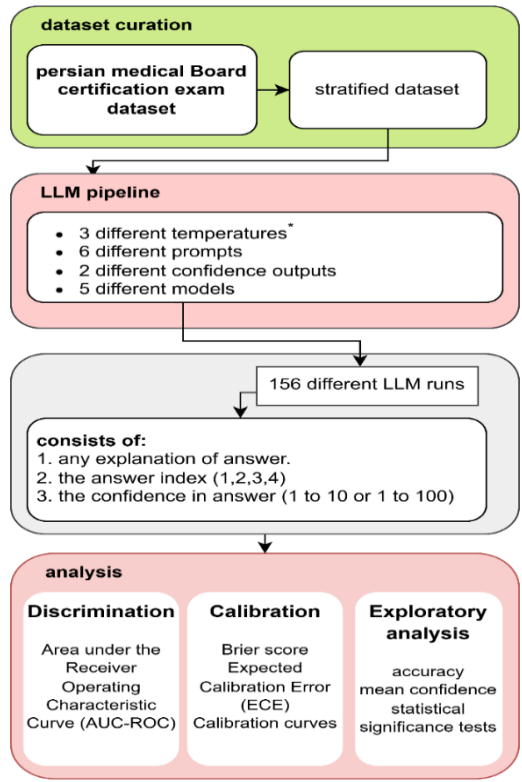

**Fig. 1.** The overall pipeline of the experiment. *The o3-mini model only supports temperature of one.

### 3.1  Data Collection and Preprocessing

The dataset was compiled from Persian specialized and fellowship board examinations administered in 2022 and 2023 across multiple medical specialties (see Figer 2). It comprises approximately 12,000 multiple-choice questions in Persian, each with an English translation verified by a medical doctor. These examinations, developed by experts using standard reference textbooks, include answer keys created by the question authors and validated by an independent M.D. Due to computational and financial constraints, a stratified random sample of 300 questions was selected, with metadata identifying each question's specialty and field. Between 5 and 8 questions per exam were chosen to ensure a robust assessment of the models' domain-specific knowledge and confidence calibration, particularly in complex, edge-case medical scenarios.

### 3.2  Experimental Setup

In our experiments, we employed the following artificial intelligence models (accessed and used on January 2025):

1- Llama 3.3-70B: Developed by Meta AI, this model comprises 70 billion parameters.[1]
2- Llama 3.1-8B: Developed by Meta AI, this model comprises 8 billion parameters.[2]
3- DeepSeek-v3: Developed by Deep Seek, this model comprises 671 billion parameters.[3]
4- GPT-4o: Developed by OpenAI (gpt-4o-2024-08-06); its exact specifications have not been publicly disclosed.[4]
5- o3-mini: Developed by OpenAI and belonging to the model reasoning family, its architecture has not been publicly disclosed.[5]

This selection was driven by an interest in examining the effects of differing model architectures from three distinct companies, variations in model size on domain-specific knowledge, and contrasts between general LLMs and those explicitly optimized for reasoning tasks. In addition, the study aimed to evaluate the robustness of these approaches when implemented in locally hosted environments—a consideration of particular importance where data privacy is critical.

The prompting techniques used in this study were designed to probe a range of configurations in both model accuracy and confidence calibration. Specifically, (1) CoT Prompting guides the LLM to undertake step-by-step reasoning; (2) Few-Shot Prompting supplies exemplar questions and corresponding answers to shape the model's expected outputs; (3) Hybrid CoT and Few-Shot Prompting combines these two methods, requiring the model to provide extensive reasoning while referencing example outputs and justifications; (4) Zero-Shot Raw Prompting presents the LLM only with the multiple-choice query and directs it to produce an answer alongside a confidence score; (5) Expert Mimicry Prompting instructs the LLM to adopt the perspective of a knowledgeable expert; and (6) Emotional Prompting highlights that the accuracy of the response and its associated confidence score may carry real-world implications for patient outcomes.

For each prompting technique, the models produced both an answer and a confidence score on two scales—ranging from 1 to 10 and 1 to 100. Moreover, three temperature settings (0.3, 0.7, and 1.0) were applied to control response variability, although OpenAI's o3-mini is limited to a temperature setting of 1.0. Supplementary Material 2 provides the full set of prompt templates.

### 3.3   Structured Output Explanation:

The Final Answer data model is implemented to standardize the output format of the experiments. It encapsulates a detailed justification that records every step outlined in the prompt except for the final answer or the confidence score, ensuring full transparency of the reasoning process. In addition, it includes a final answer field designed to hold a numerical result limited to the set {1, 2, 3, 4} and a confidence score field that quantifies the reliability of the final answer on a predefined scale, such as 1–10 or 1–100. This structure not only facilitates systematic evaluation and reproducibility of experimental outcomes but also enhances interpretability by explicitly separating the reasoning process from the conclusive metrics.

---

[1] https://huggingface.co/meta-llama/Llama-3.3-70B-Instruct
[2] https://huggingface.co/meta-llama/Llama-3.1-8B-Instruct
[3] https://huggingface.co/deepseek-ai/DeepSeek-V3
[4] https://huggingface.co/blackboxai/gpt-4o
[5] https://huggingface.co/o3/o3-mini

### 3.4    Prompt Format Explanation:

The prompts dictionary serves as a comprehensive repository of various prompt templates applied throughout the experiments. It encompasses a range of prompting strategies—including chain-of-thought (cot), few-shot, combined cot and few-shot, roleplay, zero shot, and emotional prompts—each tailored to different experimental conditions and specified by varying token limits (10, 100, and qualitative formats). This organized approach enables the systematic assessment of how different prompt configurations affect model performance, offering insights into the interplay between prompt design and the reasoning capabilities of artificial intelligence models.

### 3.5    Evaluation Metrics

Answers are considered correct only when they precisely match the verified ground truth; all other responses are marked as incorrect. Specifically, (1) Accuracy Measurement relies on a binary scoring scheme that underpins the calculation of overall accuracy across test samples. (2) Confidence Analysis determines the mean confidence for each experimental setting by averaging the confidence scores reported by the models over all test cases; these averages are independently calculated for the two scales (1–10 and 0–100). (3) Calibration Metrics gauge how closely expressed confidence levels correspond to actual performance. The AUC-ROC is computed to assess how effectively the confidence scores discriminate between correct and incorrect responses. The Brier Score measures the mean squared error between the predicted probabilities (i.e., confidence values) and the binary outcomes, while the ECE captures the disparity between the model's confidence estimates and the observed accuracy across distinct confidence bins.

### 3.6    Implementation Details and Reproducibility

All experiments were conducted using Python 3.9 within a Conda virtual environment to maintain package consistency. The OpenAI Python package (v1.58.1) served as the interface to LLMs, including those accessible through the Fireworks API, thereby standardizing all LLM operations. Asynchronous execution, implemented via the Asencio package, enhanced the operational efficiency of the LLM runs. Data handling and manipulation were performed using pandas (v2.2), ensuring structured Data Frame operations throughout.

## 4    Results

The results of each experiment (model, temperature, prompt, confidence elicitation handling) are provided in Figures 2 and 3. This consists of AUC-ROC, brier score, ECE, accuracy and mean confidence.

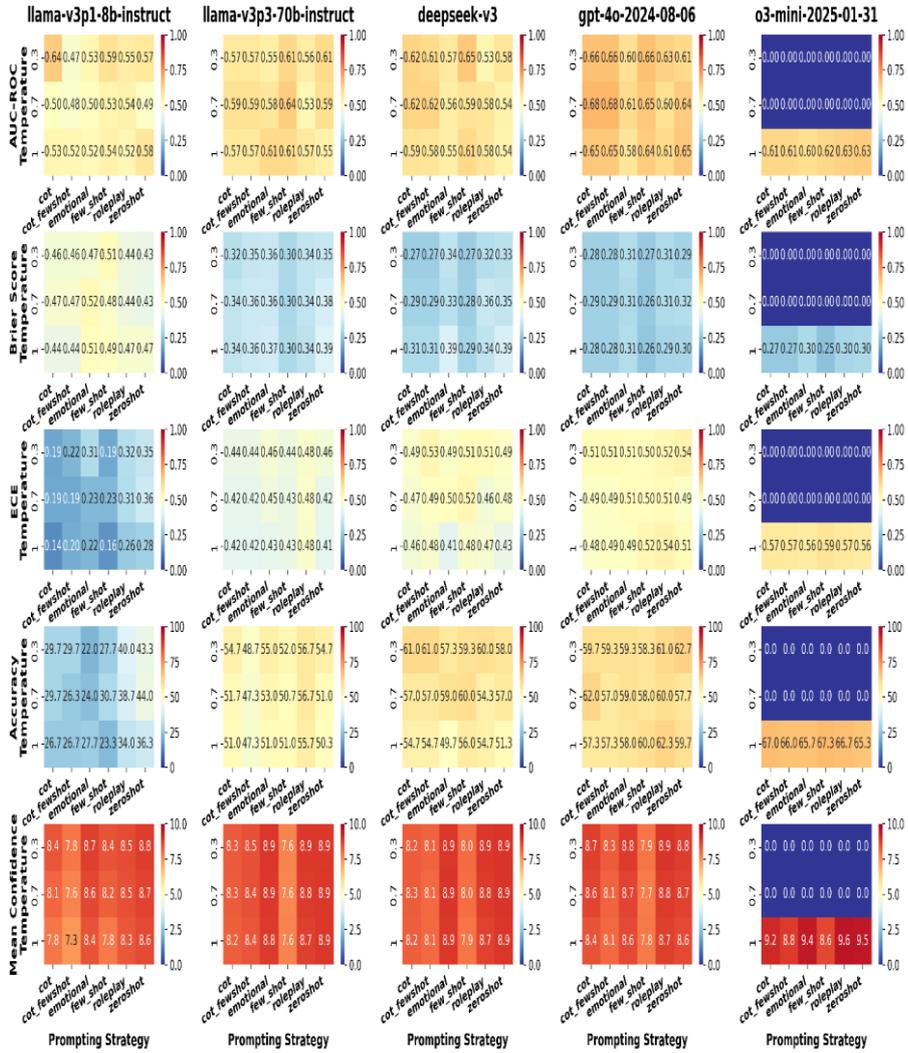

**Fig. 2** The comprehensive results of experiments utilizing five models (columns) for four evaluation metrics and the mean confidence score reported by the Large Language Model (LLM) (rows). Each graph presents a heatmap stratified by prompt (chain of thought, chain of thought with few-shot examples, emotional pressure, few-shot examples, expert role mimicry, and zero-shot simple prompt) and temperature, employing a confidence elicitation method ranging from 1 to 10.

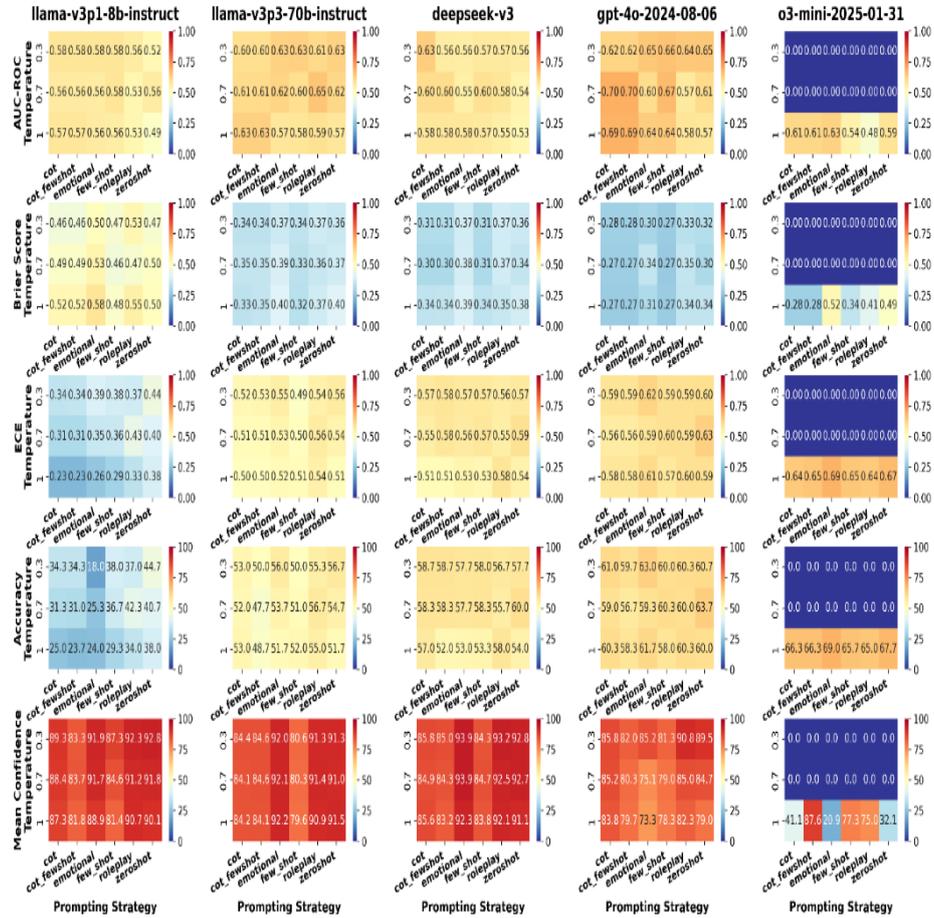

**Fig. 3** The comprehensive results of experiments utilizing five models (columns) for four evaluation metrics and the mean confidence score reported by the Large Language Model (LLM) (rows). Each graph presents a heatmap stratified by prompt (chain of thought, chain of thought with few-shot examples, emotional pressure, few-shot examples, expert role mimicry, and zero-shot simple prompt) and temperature, employing a confidence elicitation method ranging from 1 to 100.

### 4.1 Accuracy

The highest performing model was o3-mini, achieving 69.00% accuracy with emotional prompting at temperature 1.0 (1-100 confidence), 67.67% with zero shot prompting at temperature 1.0 (1-100 confidence), and 67.33% with few-shot prompting at temperature 1.0 (1-10 confidence). In contrast, Llama 3.1 8b showed the lowest performance, with 18.00% accuracy using emotional prompting at temperature 0.3 (1-100 confidence), and 22.00% accuracy for both CoT prompting at temperature 1.0 (1-10 confidence) and emotional prompting at temperature 0.3 (1-100 confidence).

Looking at the overall mean accuracies across all configurations, o3-mini achieved 66.36 ± 0.70%, followed by GPT-4o with 59.72 ± 0.58%, DeepSeek-v3 with 56.52 ± 0.85%, Llama 3.3 70b with 52.42 ± 0.91%, and Llama 3.1 8b with 31.67 ± 2.33%.

### 4.2   Discrimination

The lowest area under the AUC-ROC values were observed with the Llama-v3p1-8b-instruct model, achieving 0.474 with CoT few-shot prompting at temperature 0.3 (1–10 confidence), 0.481 with CoT prompting at temperature 0.7 (1–100 confidence), and with the o3-mini-2025-01-31 model, which obtained 0.475 using roleplay prompting at temperature 1.0 (1–100 confidence). In contrast, the highest AUC-ROC values were obtained with the gpt-4o-2024-08-06 model, reaching 0.699 with CoT few-shot prompting at temperature 0.7 (1–100 confidence), 0.689 with CoT few-shot prompting at temperature 1.0 (1–100 confidence), and 0.681 with CoT few-shot prompting at temperature 0.7 (1–10 confidence).

The mean AUC-ROC results with confidence intervals for each model are as follows. The gpt-4o-2024-08-06 model achieved a mean AUC-ROC of 0.627±0.01, while the Llama-v3p3-70b-instruct model obtained a mean AUC-ROC of 0.593±0.01. The o3-mini-2025-01-31 model recorded a mean AUC-ROC of 0.592±0.03. The DeepSeek-v3 model demonstrated a mean AUC-ROC of 0.577±0.01, and finally, the Llama-v3p1-8b-instruct model achieved a mean AUC-ROC of 0.541±0.01.

### 4.3   Calibration

The highest-performing models were o3-mini, achieving a 0.248 Brier score with few-shot prompting at temperature 1.0 (1–10 confidence), GPT-4o achieving 0.264 with few-shot prompting at temperature 1.0 (1–10 confidence), and GPT-4o also achieving 0.264667 with CoT prompting at temperature 1.0 (1–10 confidence). In contrast, the models with the highest (worst) Brier scores were Llama 3.1 8b, with 0.580 using roleplay prompting at temperature 1.0 (1–100 confidence), Llama 3.1 8b with 0.582 using few-shot prompting at temperature 1.0 (1–10 confidence), and Llama 3.1 8b with 0.599 using emotional prompting at temperature 1.0 (1–100 confidence).

### 4.4   Impact of Model, Prompt, and Temperature

Figures 4, 5, and 6 illustrate the varying impact of model, prompt, and temperature on performance, respectively. On average, GPT-4o exhibited the lowest mean Brier score of 0.30 ± 0.01, followed by DeepSeek-v3 with 0.33 ± 0.01 and gpt o3 mini with 0.35 ± 0.05. Similarly, Llama 3.3 70b attained a mean Brier score of 0.35 ± 0.01, while Llama 3.1 8b demonstrated the highest average Brier score of 0.49 ± 0.01, indicating greater overall uncertainty. These findings further underscore the varying degrees of confidence calibration across different models.

As illustrated in Figure 7, the highest ECE values were observed with the o3-mini model, reaching 0.687 with emotional prompting at temperature 1.0 (1–100 confidence), 0.673 with zero-shot prompting at temperature 1.0 (1–100 confidence), and 0.654 with Chain of Thought (CoT) few-shot prompting at temperature 1.0 (1–100 confidence). Conversely, the lowest ECE values were achieved with the Llama 3.1 8b model, with 0.142 using CoT prompting at temperature 1.0 (1–10 confidence), 0.156 using few-shot prompting at temperature 1.0 (1–10 confidence), and 0.190 using CoT prompting at temperature 0.3 (1–10 confidence).

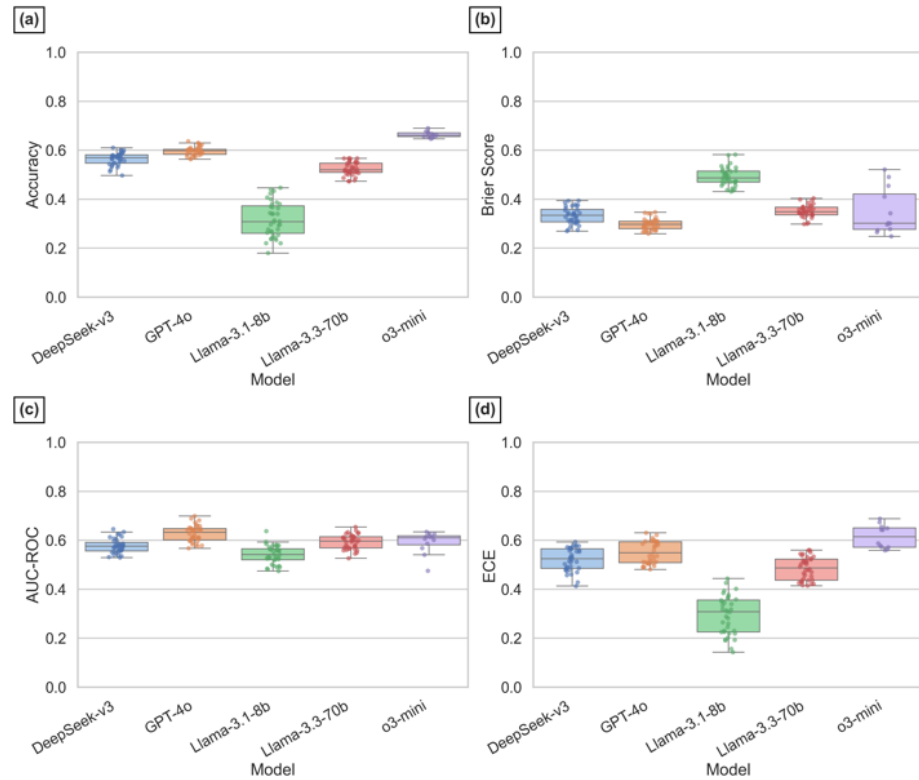

**Fig. 4.** The performance of LLMs on accuracy (a), Brier score (b), AUC-ROC (c), and ECE metrics stratified by five distinct models.

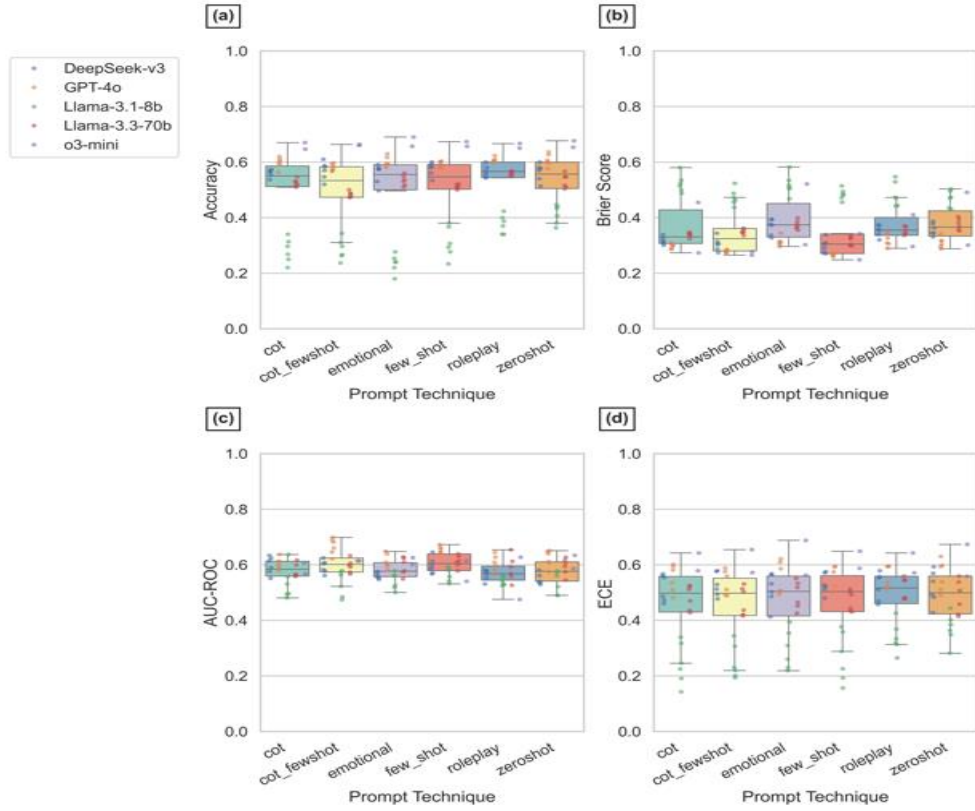

**Fig. 5.** The impact of prompt engineering utilizing six techniques: chain of thought (CoT), chain of thought combined with few-shot examples (CoT_fewshot), emotional pressure (emotional), few-shot examples (few_shot), expert role mimicry (roleplay), and zero-shot simple prompt (zeroshot) on four metrics: accuracy (a), Brier score (b), AUC-ROC (c), and ECE. Data points for each model are differentiated by color.

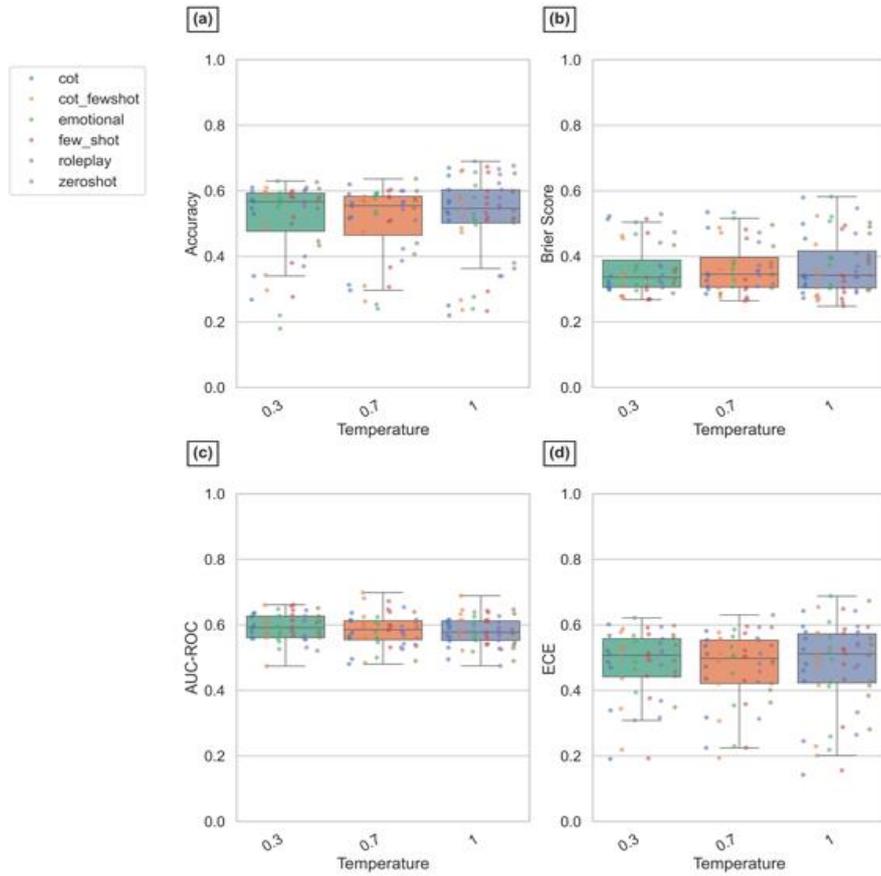

**Fig. 6.** The impact of varying temperatures on four metrics: accuracy (a), Brier score (b), AUC-ROC (c), and ECE. Data points for each model are differentiated by color. Data points for each prompt are represented by six techniques: chain of thought (CoT), chain of thought combined with few-shot examples (CoT_fewshot), *emotional* pressure (emotional), few-shot examples (few_shot), expert role mimicry (roleplay), and zero-shot simple prompt (zeroshot).

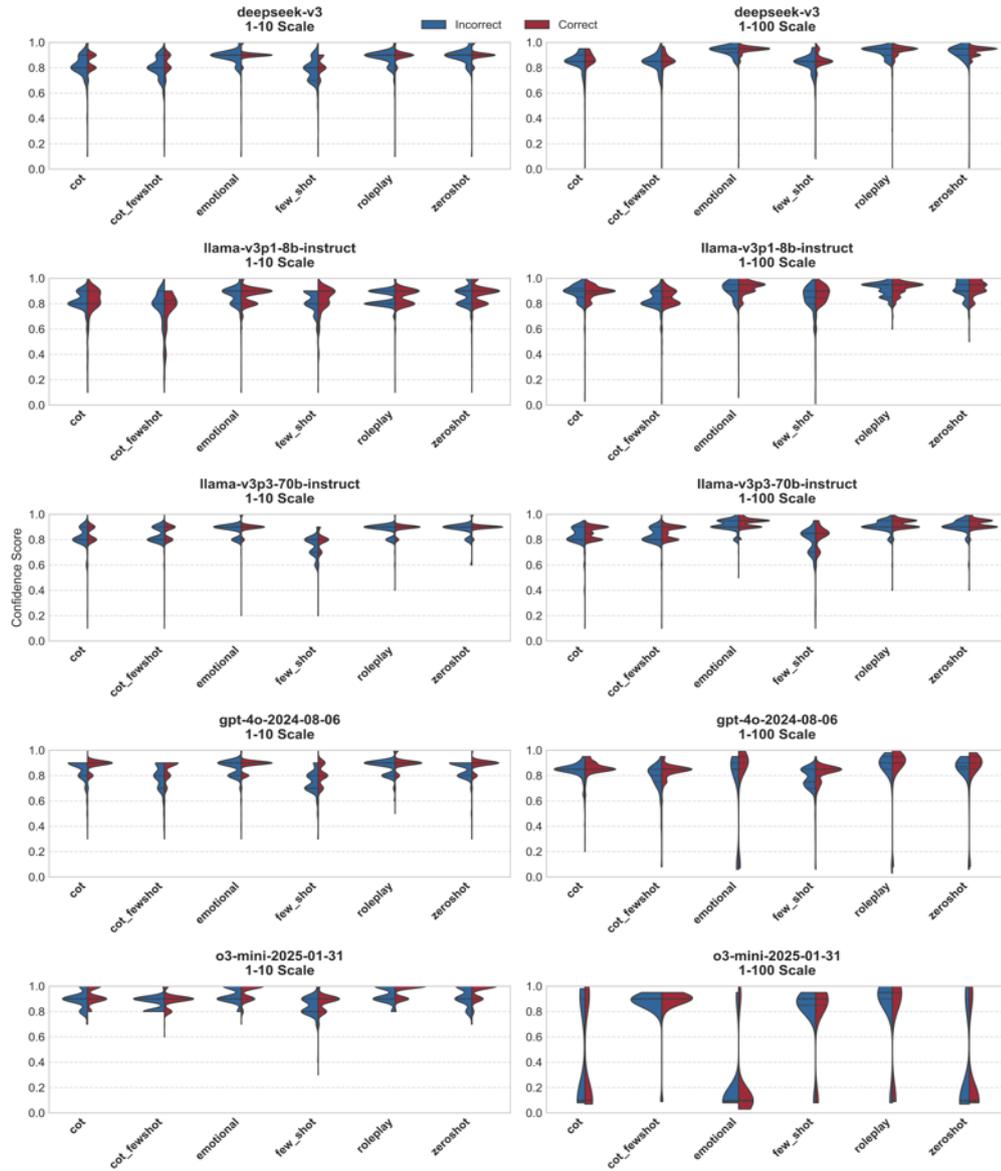

**Fig. 7.** The confidence scores reported by Large Language Models (LLMs) stratified by the correctness of LLM responses to the question (correct: blue; incorrect: red). The right column illustrates the confidence elicitation scoring of 0-100, while the left column depicts the confidence elicitation scoring of 0-10 for various models.

## 5  Discussion

The integration of reasoning models that generate intermediate rationales prior to final outputs has significantly improved the accuracy of LLMs in medical applications, especially in complex or edge-case scenarios. This trend is corroborated by our results, where O3-mini achieved the highest accuracy across all 156 evaluated model configurations. However, this performance gain coincides with increased overconfidence in self-assessed correctness compared to its predecessor, GPT-4o, which exhibited superior calibration despite lower accuracy. Notably, this pattern extends to standard models employing CoT prompting, where accuracy improvements are systematically accompanied by inflated confidence estimates. Open-source architectures such as Llama-3.3-70B and DeepSeek-V3 demonstrate promising capabilities for privacy-sensitive medical applications, though they underperform relative to state-of-the-art proprietary models. Conversely, the persistent inadequacy of smaller models like Llama-3.1-8B—even when optimized for edge deployment—underscores their current unsuitability for high-stakes clinical use. This underscores the critical need for robust architectures when addressing complex clinical queries.

A critical finding across all evaluated models is their systemic overconfidence in self-evaluated correctness. No model achieved a Brier score below 0.25 (commonly approximating random-chance performance in a four-option setting) or an AUC-ROC score exceeding 0.7 (a minimal threshold often cited for acceptable discriminatory power). These results collectively indicate that current confidence elicitation methods remain unreliable for gauging LLM knowledge and uncertainty in medical domains, particularly in high-stakes contexts where misjudged confidence could have serious repercussions.

Prompting strategies exhibited divergent effects on model performance. Emotional prompting—framing queries through narratives emphasizing patient vulnerability or caregiver distress—consistently inflated confidence scores across architectures. This phenomenon, analogous to a psychological mechanism that prioritizes reassurance over accuracy, poses critical risks in clinical settings where overconfidence can compromise patient safety. Crucially, this artificial confidence inflation degraded both calibration (Brier score) and discrimination (AUC-ROC) metrics, suggesting that emotionally charged prompts may impair objective clinical judgment. In contrast, expert mimicry prompting increased absolute accuracy despite similarly exacerbating overconfidence, whereas emotional prompting showed no statistically significant improvement over zero-shot baselines.

CoT and few-shot prompting also yielded nuanced performance patterns. While CoT universally improved accuracy, it simultaneously amplified confidence, particularly in larger models such as GPT-4o and Llama-3.3-70B when compared to few-shot or combined few-shot+CoT approaches. This suggests that explicit reasoning steps, while beneficial for problem-solving, may inadvertently reinforce confirmation bias through self-validating rationales. Few-shot prompting produced more balanced gains in accuracy and calibration—most noticeably under conservative temperature settings (0.3–0.7)—indicating that concrete examples can help align confidence estimates with empirical evidence.

Temperature tuning revealed model-specific sensitivities in calibration dynamics. Though typically employed to adjust output variability, changes in temperature did not uniformly improve ECE or discrimination metrics. GPT-4o exhibited relatively stable calibration across temperature variations compared to O3-mini, implying that some architectures inherently constrain confidence-expression patterns, making them less susceptible to sampling-based heuristics.

The disparity in confidence reporting between 1–10 and 1–100 scales raises additional methodological concerns for self-assessment elicitation. Contrary to the assumption that finer-grained scales provide more nuanced confidence estimates, several models merely inflated their scores without improving calibration. These findings indicate that more granular confidence scales do not automatically yield better alignment with actual performance, highlighting the need for structured post-hoc calibration techniques [4,12].

From a deployment perspective, the performance gap between open-source and proprietary models illustrates key trade-offs in privacy-sensitive environments. While Llama-3.3-70B and DeepSeek-V3 did not match the accuracy of GPT-based systems, their ongoing improvements and flexible deployment options suggest they could evolve into clinically viable solutions for data-sensitive contexts. However, the pronounced underperformance of smaller models like Llama-3.1-8B—even with edge optimization—confirms that resource-constrained architectures currently lack the capacity to internalize complex medical knowledge while maintaining consistent calibration.

These findings collectively demonstrate that accuracy enhancements in medical LLMs do not inherently translate into reliable uncertainty assessments. While prompting strategies such as CoT variants boost correctness, they risk amplifying confidence in erroneous outputs. This necessitates a dual optimization framework that combines domain-specific prompt engineering with rigorous post-hoc calibration to mitigate risks in clinical applications. For high-stakes medical settings, such an integrated approach is vital to prevent the harm that can arise from overly confident yet inaccurate model responses.

## 6    Conclusion

The findings of this study underscore the critical role of prompt engineering in shaping both the accuracy and confidence elicitation of LLM outputs within high-stakes medical contexts. Despite notable gains in performance using advanced prompting techniques—such as CoT and Few-Shot methods—overconfidence consistently emerges as a systemic challenge, highlighting the importance of a dual strategy that combines structured prompt designs with robust post-hoc calibration.

From a philosophical standpoint, this work resonates with the broader discourse on reflective AI [18][19], wherein models are encouraged to exhibit epistemic humility—a core principle in both classical epistemology and contemporary ethics. The outcomes reveal that achieving reliable AI systems is not merely a technical hurdle but also a conceptual one, demanding careful reflection on the nature of knowledge, uncertainty, and trust. In this sense, prompt engineering becomes a catalyst for embedding philosophical rigor into AI design, ultimately fostering more transparent, accountable, and ethically aligned intelligent systems.

**Acknowledgment.** This research was undertaken, in part, thanks to funding from the Canada Research Chairs Program.

GitHub repository,
https://github.com/narimannr2x/confidence_elicitation_different_prompts